# Synthetic Graphene Grown by Chemical Vapor Deposition on Copper Foils

TING FUNG CHUNG,[1,3] TIAN SHEN,[1] HELIN CAO,[1,3] LUIS A. JAUREGUI,[2,3] WEI WU,[4] QINGKAI YU,[5] DAVID NEWELL[6] AND YONG P. CHEN[1,2,3]

[1]*Department of Physics,* [2]*School of Electrical and Computer Engineering,* [3]*Birck Nanotechnology Center, Purdue University, West Lafayette, Indiana 47907, USA*

[4]*Department of Electrical and Computer Engineering, University of Houston, Houston, Texas 77204, USA*

[5]*Ingram School of Engineering, and Materials Science, Engineering and Commercialization Program, Texas State University, San Marcos, Texas 78666, USA*

[6]*Physical Measurement Laboratory, National Institute of Standards and Technology, Gaithersburg, Maryland 20899, USA*

The discovery of graphene, a single layer of covalently bonded carbon atoms, has attracted intense interests. Initial studies using mechanically exfoliated graphene unveiled its remarkable electronic, mechanical and thermal properties. There has been a growing need and rapid development in large-area deposition of graphene film and its applications. Chemical vapour deposition on copper has emerged as one of the most promising methods in obtaining large-scale graphene films with quality comparable to exfoliated graphene. In this chapter, we review the synthesis and characterizations of graphene grown on copper foil substrates by atmospheric pressure chemical vapour deposition. We also discuss potential applications of such large scale synthetic graphene.

## 1. Introduction

Graphene, the first two-dimensional atomic crystal, shows exceptional electronic[1,2] and thermal properties[3], robust mechanical strength[4], unique optical[5], other physical properties, etc. Systematical investigations in the physical properties of graphene began in the mechanical exfoliation graphene from graphite. Amazingly, mechanical exfoliation gives highly crystalline graphene flakes, showing high carrier mobility of ~10 000 $cm^2$/Vs on a Si wafer and >~100 000 $cm^2$/Vs when suspended or deposited on hexagonal boron nitride (h-BN),even at or close to room temperature (RT)[2,6,7]. However, its applications are limited by the small flake size and non-uniformity in the number of graphene layers in the exfoliated flakes from graphite. There are several methods to synthesize graphene films such as thermal decomposition of silicon carbide (SiC)[8,9], chemical reduction of graphene oxide (GO) film[10,11], and metal catalytic chemical vapor deposition (CVD) growth[12,13]. Table 1 summarizes the maximal reported sample size and RT charge carrier mobility of graphene made by as-mentioned methods. High mobility (~10 000 $cm^2$/Vs)[14] epitaxial graphene can be obtained in thermal decomposition of SiC, however high cost and limited SiC wafer size may restrain its wide applications. The chemical reduction of GO can also produce graphene-based connected films in large-scale, but the major drawback is low electrical mobility (~1 $cm^2$/Vs)[15,16] originated from their defective structures. Among these methods, metal catalytic CVD has become one of the most promising ways in synthesizing large-scale graphene films since this method gives transferable high-quality graphene films with high yield, relatively low cost and large area whose size is limited only by the metal substrate and furnace. The catalytic growth of multilayer graphene on metals can be traced back to 1939[17], even before the first report of the success of obtaining single-layer graphene (SLG) by mechanical exfoliation. In recent CVD growth, various metals, such as Ni[18], Cu[12], Ni-Cu alloy[19,20], Co[21], Ir[22], Ru[23], and Pd[24], have been used for graphene growth. Particularly, Cu has become the most widely used because the low carbon solubility of Cu facilitates a large-area, uniform growth of single-layer graphene. Moreover, the availability of large, inexpensive Cu foil substrates suits the development of graphene-based applications.

Table 1. Maximal reported sample size and room temperature (RT) charge carrier mobility of graphene synthesized by different methods.

| Graphene production method | Max. sample size (mm) | RT charge carrier mobility achieved (cm$^2$/Vs) | Ref. |
|---|---|---|---|
| Mechanical exfoliation | ~1 | ~1 x 10$^5$ | 7 |
| CVD on Cu | ~1000 | 10 000 | 13,25 |
| Epitaxial growth on SiC | ~100 | 10 000 | 8,9 |
| Graphite oxide reduction | ~1000 | ~1 | 15,16 |

Owing to the growth kinetics of graphene on typical Cu foil substrates, the large-scale

SLG grown on Cu foil shows polycrystallinity with domain boundaries[25-27]. The presence of domain boundaries in graphene can limit its physical properties compared to that of mechanically exfoliated graphene (typically single crystalline). For instance, CVD-grown graphene usually shows a lower mobility, ranging from several hundreds to ~5000 cm$^2$/Vs, and domain boundaries are considered as one of the important causes. Despite the polycrystalline nature and some degree of non-uniformity of graphene film grown on Cu, the material still demonstrates ambipolar field-effect, high quality 2D electron gas quantum Hall effect (QHE)[13,28,29], similar to mechanically exfoliated graphene. Here, we review the synthesis and properties of CVD-grown graphene, demonstrated using examples primarily from our work in recent years. Particularly, we will review the growth of CVD graphene on Cu foils using atmospheric pressure (AP) CVD, and the transfer of CVD grown graphene films on arbitrary substrates, the Raman characterization, the electronic transport in transferred CVD graphene, as well as some application prospects of CVD graphene.

## 2. Atmospheric pressure CVD grown graphene films

The growth recipes of CVD graphene can vary between different groups and growth setups. Briefly, they are classified into two main categories based on the working pressure: Low-pressure (LP) CVD and atmospheric pressure (AP) CVD. The working pressures for graphene growth at LPCVD and APCVD are ~ 0.1 – 1 Torr and ~760 Torr[12,13,25], respectively. The kinetics of the growth at LP and AP are different, leading to a variation in the shape, size and uniformity of graphene domains. For instance, the typical shape of graphene domains grown in LPCVD is lobe-flower-shape[30], whereas hexagonal shape of graphene domains is usually obtained in APCVD[25]. In the recent literatures of CVD grown graphene, a range of working pressures between 10 to 760 Torr and various ratios between carbon precursor and hydrogen gas have been explored[31,32], increasing the graphene single crystal domain size up to millimeter scale. Table 2 summarizes a collection of growth conditions of several examples of CVD graphene grown on Cu foils, the average size of graphene domain, and field-effect (FE) charge carrier mobility measured at RT unless stated otherwise. It is noted that charge carrier mobility depends on the mean size of graphene domains, influenced by the growth condition. And the carrier mobility of CVD grown graphene is approaching to that of exfoliated graphene.

Table 2. Collection of various growth parameters (working pressure, growth temperature, flow rate of methane ($CH_4$) and $H_2$), and sample characteristics (size of single crystal graphene domains and FE charge carrier mobility measured at RT unless stated otherwise) of CVD grown graphene on Cu foils. Values drawn from previous published references and another recent example ("this work") from our work.

| Working | Growth | $CH_4$ (sccm) | $H_2$ (sccm) | Average | FE carrier | Ref. |
|---|---|---|---|---|---|---|

| pressure (Torr) | temp. (°C) | | | domain size (μm) | mobility at RT (cm²/Vs) | |
| --- | --- | --- | --- | --- | --- | --- |
| 0.5 | 1000 | 35 | 2 | ~10-20 | NA | 12 |
| 0.16 – 0.46 | 1035 | 7 - 35 | 2 | ~30 | ~15 000 | 30 |
| 0.04 | 1035 | 0.5 – 1.3 | 2 | ~400 | ~4000 | 33 |
| 760 | 1050 | NA | 310[a] | NA | ~2500[b] | 28 |
| 760 | 1050 | 300[c] | 10 | ~10 | ~10 000[d] | 25, 34 |
| 760 | 1050 | 40[e] | 460[f] | ~10 | ~5000 | this work |
| 10 | 1045 | 0.5 | 500 | ~400 | NA | 35 |
| 108 | 1077 | 0.15 | 70 | ~2000 | ~11 000[g] | 32 |

[a] Total gas flow rate is 310 sccm (70 ppm $CH_4$, $H_2$/Ar = 1:30).
[b] The FE mobility was measured at ~0.6 K.
[c] 8-50 ppm concentration in $CH_4$.
[d] The FE mobility was measured on a single crystal domain at ~4 K.
[e] 500 ppm concentration in $CH_4$.
[f] 460 sccm of 5% $H_2$ balanced in Ar.
[g] The FE mobility was measured on a single crystal domain at ~300 K.

A typical APCVD system to grow graphene is shown in Fig. 1a. The growth substrate used is Cu foil (99.8% purity, Alfa Aesar). A typical growth procedure (used for the graphene samples in most of the examples described in this review, note moderate adjustment of parameters are often made for different growth and different CVD systems) is as follows. A 25-μm thick Cu foil substrate was cleaned by acetone and isopropanol (IPA) followed by acetic acid to remove native oxide. The cleaned Cu foil was thoroughly dried by a nitrogen gas and then loaded into the APCVD system. The reaction chamber was evacuated to ~20 mTorr, and then filled back to ambient pressure with a forming gas (5% $H_2$/Ar). After this, the temperature was increased to 1050 °C with flowing forming gas of 460 sccm. The Cu foil was annealed for 30 min.

Then the graphene growth was performed by flowing methane (500 ppm $CH_4$ diluted by Ar) for 120 min. After the growth, the $CH_4$ flow was turned off and the Cu foil was cooled down naturally.

Transferring as-grown graphene film from the Cu substrate to an insulating substrate is a critical step for fabricating electronic devices. PMMA assisted transfer technique is commonly applied because of its simplicity and repeatability[12]. In a typical transfer, a graphene film on Cu substrate was first coated with PMMA (950PMMA-A4, MicroChem) by spin-coating, then slightly dried on a hotplate. The graphene on the reverse side (not covered by PMMA) of Cu was removed by plasma etching. The PMMA-graphene-Cu stack was floated on a copper etchant (0.25 g/mL $FeCl_3$ in water) overnight. After copper etching, the PMMA-graphene membrane (shown in Fig. 1b) was scooped out and transferred to several baths of DI water and SC solutions for rinsing[36]. It was then scooped out again with a target $Si/SiO_2$ substrate and dried in air overnight before immersion in a bath of acetone to dissolve the PMMA, followed by rinsing in IPA and drying with nitrogen gas flow. Fig. 1(a) displays a three layer stacked graphene on a cover glass made by layer-by-layer transfer. Optical contrast can be used to distinguish the difference in the number of graphene layers. Fig. 1(b) shows an optical image of a predominant monolayer graphene film grown by CVD then transferred on a Si substrate with ~300 nm oxide.

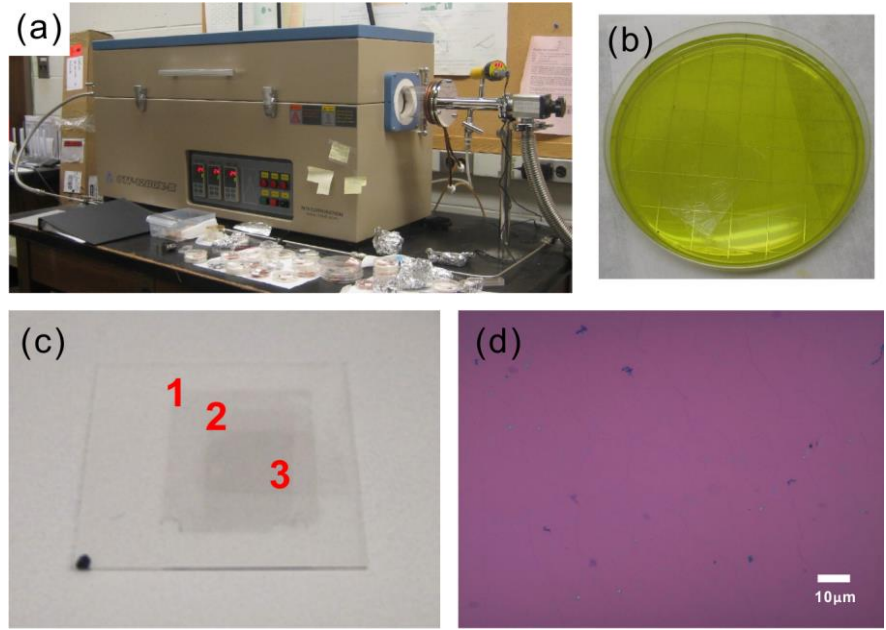

Fig. 1. Growth and transfer of CVD graphene film. (a) Photograph of a tube furnace CVD system for graphene growth at Purdue University. (b) Transparent PMMA/graphene membrane floating on copper etchant. (c) 3 layers of stacked CVD graphene on a cover glass made by consecutively transferring 3 graphene films. Optical contrast of the stacked graphene illustrates discernable difference in the number of layers. (d) Optical image of a single-layer graphene film transferred on a Si wafer with 300 nm thermal oxide.

**3. Structural and morphological characterizations by Raman and atomic force microscopy (AFM)**

Raman spectroscopy is a swift and non-destructive method to characterize the crystal quality, number of layers, and doping level of graphene film through exciting phonon vibrational modes in graphene and probing electron-phonon interactions[37-39]. In the examples shown here, micro-Raman spectra were obtained on a transferred CVD graphene onto a Si wafer with 300 nm thermal grown oxide using a Horiba Jobin Yvon Xplora confocal Raman microscope. Careful analysis of Raman peaks confirms the presence of SLG and the success in graphene transfer. Fig. 2(a) presents a representative Raman spectrum of the transferred CVD graphene film using a 532 nm excitation laser. The prominent features of SLG are G peak at ~ 1580 cm$^{-1}$ and a symmetric 2D peak at ~ 2700 cm$^{-1}$ with FWHM of ~32 cm$^{-1}$. The insignificant D peak in the spectrum near ~ 1350 cm$^{-1}$ indicates the high quality and low defects. In general, the appearance of the D peak signifies disorder in the carbon lattice such as the edge of domain and domain boundaries[25], and lattice defects/distortion[40], etc. In addition to the line shape of 2D peak, it is known that the ratio of $I_{2D}/I_G$ can be used to distinguish the number of graphene layers[11]. The typical $I_{2D}/I_G$ ratio of single layer and bilayer exfoliated graphene is ~2-3 and slightly less than 1, respectively[37]. For our transferred CVD graphene shown in Fig. 2(a), the $I_{2D}/I_G$ dominantly has a ratio of 2-3, similar to that measured on exfoliated SLG. The large $I_{2D}/I_G$ ratio sometimes measured in transferred CVD graphene is speculated to be related to the slightly suspend graphene film during transfer process[41,42]. The inset of Fig. 2(a) displays the ratio of $I_{2D}/I_G$ and full-width hall maximum (FWHM) of both G and 2D peaks of typical exfoliated SLG and our CVD-grown SLG. Based on both $I_{2D}/I_G$ and FWHM of 2D peak, the graphene film is dominantly SLG. Figure 2(b) and (c) show a representative 10 × 10 μm$^2$

Raman map of $I_D/I_G$ and $I_{2D}/I_G$, respectively, indicating the high quality and uniformity of graphene films grown by APCVD method on Cu foils.

In addition to optical and Raman characterizations, atomic force microscopy (AFM) is a versatile method to examine the thickness, number of layers and surface morphology of graphene films. Figure 2(d) shows an AFM image of a CVD grown graphene after its transfer onto a Si/SiO$_2$ substrate. Micron-sized wrinkles and small amount of particles are found on the surface of graphene. The thickness of the graphene was measured at an edge and found to be ~1.5 – 2 nm, which deviates from the expected thickness of graphene (0.35 nm). This apparent discrepancy is attributed to adsorbed molecules between the graphene and the SiO$_2$ substrate, wrinkles introduced during transfer as well as the instrument offset due to tip-substrate interaction[43].

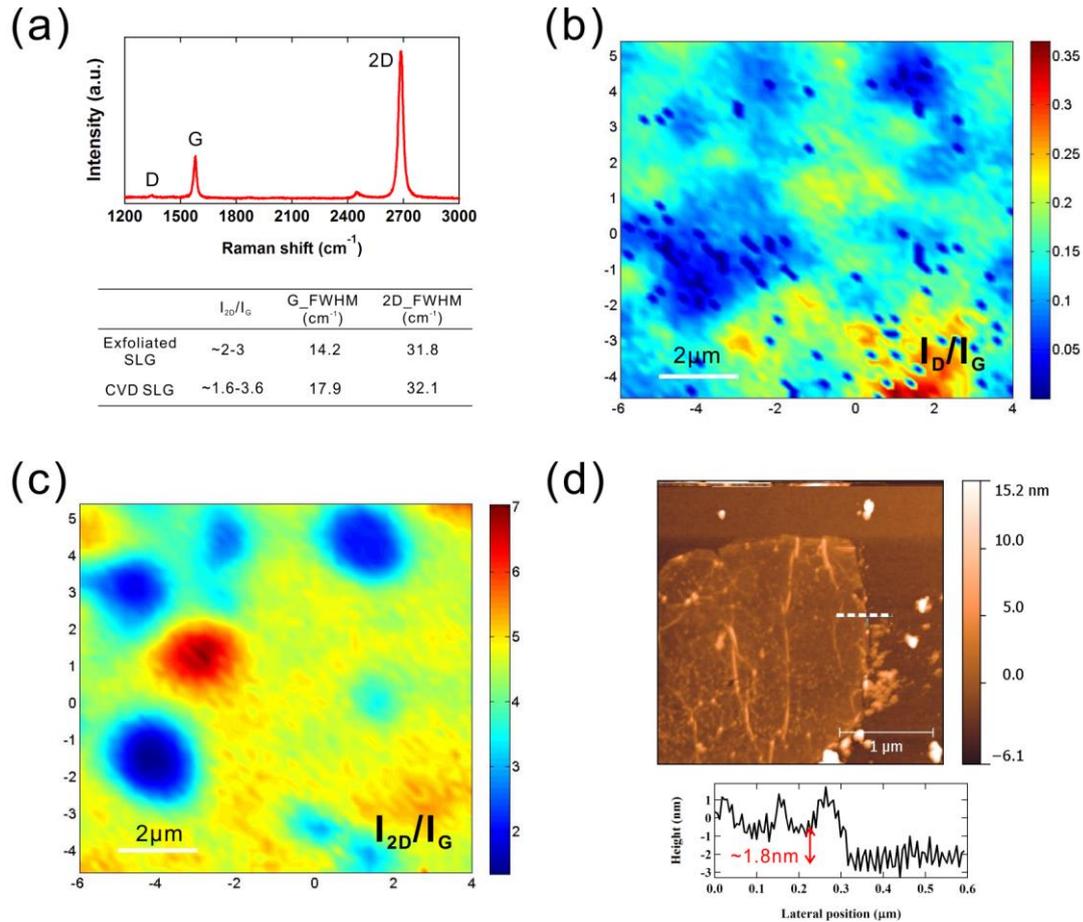

Fig. 2. Characterizations of transferred CVD graphene film on Si substrate with ~300 nm thermal oxide by Raman spectroscopy and atomic force microscopy (AFM). (a) A representative Raman spectrum of CVD single-layer graphene measured in ambient using a 532 nm excitation laser. Inset: The $I_{2D}/I_G$, the G-band FWHM and 2D-band FWHM for exfoliated SLG[44] and CVD grown SLG. (b) Representative Raman mapping of $I_D/I_G$ over a 10 × 10 μm$^2$ area. (c) Raman mapping of $I_{2D}/I_G$ of a 10 × 10 μm$^2$ area, most (~>80%) of which can be associated with single-layer graphene ($I_{2D}/I_G$ >2). (d) AFM height image and profile of a transferred CVD graphene film. The height profile is recorded along the white dashed line indicated in the AFM image.

In addition to single-layer graphene, often bilayer graphene domains are also found on CVD grown sample, shown in Fig. 3(a). There is a technological interest in growing Bernal-stacked bilayer graphene (AB stacked BLG), which has an electrically tunable band gap[45]. The growth of AB stacked BLG is relatively less studied compared to that of SLG[46-48]. Often twisted bilayer graphene (tBLG) domains (the second graphene layer is randomly rotated with respect to the first layer) can also be found in CVD grown graphene. The properties of those tBLG are determined by the relative orientations and interactions between the two graphene layers[49,50]. Micro-Raman can be utilized to characterize those BLG domains. An example of such characterizations measured on 4 twisted bilayer domains in ambient condition using 532 nm excitation wavelength (2.33 eV) is shown in Fig. 3(b). Substantial variation in $I_{2D}/I_G$ (1.5 - ~8) and FWHM of 2D peak (26 – 42 cm$^{-1}$), shown in Fig. 3(c) are observed. The data evidently show different spectral features from these twisted bilayer graphene compared to SLG and AB stacked BLG[49,50]. As illustrated in Fig. 3(a), the color contrast of SLG and BLG is apparently different when the graphene film was transferred onto a Si/SiO$_2$ substrate. The 2D band of the twisted BLG (#1, 2 and 4 shown in Fig. 3(b) and (c)) with high rotation angle (>15º) is more symmetrical and stronger ($I_{2D}/I_G$ ratio) than that of a typical 2D lineshape of SLG and AB stacked BLG (as-symmetrical lineshape with 4 Lorentzian sub-bands)[51]. As for the twisted BLG (#3), the $I_{2D}/I_G$ ratio of its spectrum is slightly larger that of AB stacked BLG ($I_{2D}/I_G$ ~1) indicating a small rotation angle between the two layers. The coupling between graphene layers depends on the rotation angle between layers and results in rotation angle dependence of the electronic properties in tBLG. CVD grown graphene provides an easy way to obtain tBLG with different rotation angles that may be interesting for studying BLG with tunable electronic structures via stacking.

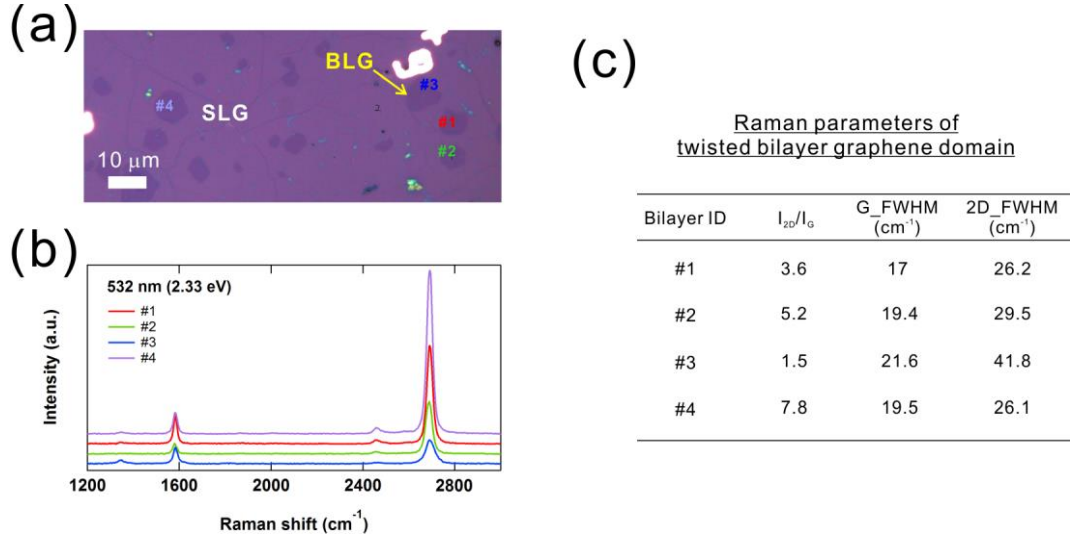

Fig. 3. Raman spectra of twisted bilayer graphene domains on Si/SiO$_2$ substrate. (a) Optical image of transferred CVD graphene film with randomly distributed bilayer graphene domains. Positions #1, #2, #3, and #4 are labeled as Raman collection spots. (b) Raman spectra of different twisted bilayer domains measured in ambient using a 532 nm excitation laser. (c) The $I_{2D}/I_G$, the G-band FWHM and 2D-band FWHM for several twisted bilayer domains shown in Fig. 3 (b).

## 4. Electronic transport properties of transferred CVD graphene

Transferred CVD graphene samples are often fabricated into Hall bar devices on Si/SiO$_2$ substrate to characterize charge carrier mobility, quantum Hall effect (QHE) and other transport properties. An example of a Hall bar device with channel length and width of 150 μm and 10 μm, respectively, is shown in the inset of Fig. 4(a). This

Hall bar device was fabricated using photolithography with e-beam evaporated metal (Ti/Au) contacts. The sample was then promptly cooled down in a variable temperature $^4$He cryostat (1.6 K to 300 K) to minimize the exposure to atmosphere, which introduces hole doping, thereby up-shifting Dirac point voltage $V_{Dirac}$. Magneto-transport measurements were performed using the low frequency ac lock-in technique with a source-drain input current of 100 nA for characterizing of the device performance. The carrier density was tuned by a back-gate voltage $V_g$ with a 300 nm thermal grown $SiO_2$ as the gate dielectric.

## 4.1 Ambipolar field-effect and carrier mobility in CVD-grown graphene devices

SLG is a gapless semi-metal with Dirac cones in the band structure[2]. When the Fermi energy ($E_F$) is close to the Dirac point (charge neutral point), which connects the upper and lower Dirac cones, the electrostatic field effect modulation of the charge carrier concentration and conduction properties is very effective. By changing the gate voltage ($V_g$), electrostatic charge carriers are induced in graphene, thereby shifting the $E_F$ either up or down. When $E_F$ is above (below) Dirac point, the dominant carriers are electrons (holes), and graphene is said to be n (p)-type. When $E_F$ is at the Dirac point, the graphene is charge neutral, exhibiting a minimal conductivity. Fig. 4(a) shows the representative four terminal longitudinal resistance $R_{xx}$ as a function of $V_g$ measured at 296 K and 1.6 K without magnetic field. It shows ambipolar FE with resistance modulation ratio of 6 and greater than 8 at 296 K and 1.6 K, respectively. Hole (electron) predominant charge transport is on the left (right) side of the peak of resistance. RT measurement shows that the $V_{Dirac}$ in the device is situated at -1.4 V, indicating a low extrinsic charge doping level. However, the $V_{Dirac}$ in the device shifts to 4.5 V after cooling down. The contaminations are probably introduced during common fabrication processes since graphene is very sensitive to charge perturbation and scattering by nearby particles on its surface. A total of 34 devices were measured in order to examine the overall electronic performance of our CVD grown graphene films. The histograms of $V_{Dirac}$ and FE carrier mobility of electron $(\mu_n)$ and hole $(\mu_p)$ measured at RT are illustrated in Fig. 4(b) and 4(c), respectively. The average $V_{Dirac}$ among 34 devices is around -2 V, indicating a low level of n-type doping. The FE carrier mobility of our graphene devices is ~5000 cm$^2$/Vs. We also measured $R_{xx}$ and $R_{xy}$ as functions of magnetic field B at fixed $V_g$ = -5 V. The Hall mobilities extracted from such measurements are found to be comparable with FET mobilities extracted from the FE measurements. The variation of charge carrier mobility is possibly attributed to both intrinsic and extrinsic disorders including domain boundaries, wrinkles, structural defects, and transfer induced impurities.

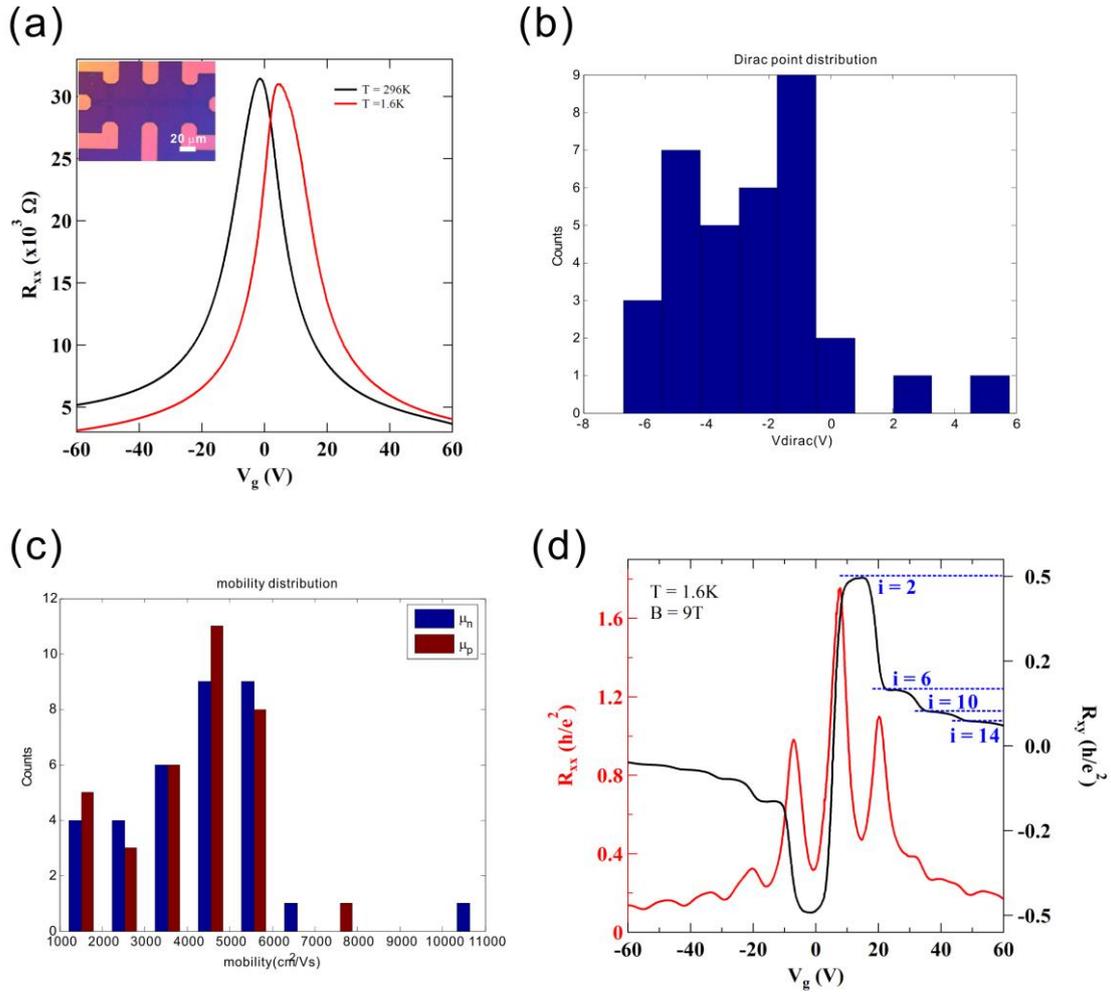

Fig. 4. Electronic transport properties of CVD graphene transferred on Si/SiO$_2$ substrate. (a) Typical ambipolar transport characteristics in resistance (R$_{xx}$) versus gate voltage (V$_g$) of back-gated CVD graphene field-effect transistor (FET) at 296 K and 1.6 K. Inset: Optical image of a typical CVD graphene Hall bar device with channel width and length of 10 μm and 150 μm, respectively. Histogram of (b) Dirac point voltage, V$_{Dirac}$, and (c) FE carrier mobility measured at room temperature in multiple devices. Notation $\mu_n$ and $\mu_p$ represent the FE mobility of electron and hole, respectively. (d) Half-integer quantum Hall effect (QHE). Hall resistance R$_{xy}$ and longitudinal resistance R$_{xx}$ as a function of V$_g$ at B = 9 T and Temp. = 1.6 K.

## 4.2 Quantum Hall effect of CVD-grown graphene

To further characterize the graphene film quality and characteristic of QHE, we have measured R$_{xx}$ and R$_{xy}$ versus V$_g$ at 1.6 K with a fixed B (perpendicular magnetic field)

= 9 T, as shown in Fig. 4(d). CVD grown graphene films possess good electronic properties can show QHE[13,29]. Unlike other 2D electron gas (2DEG) systems, the quantized condition in graphene is shifted by a half-integer which can be ascribed to Berry's phase π, implying the existence of Dirac Fermion in graphene[2,52]. The sign of reversal of $R_{xy}$ at around $V_{Dirac}$ is consistent with the ambipolar FE. Remarkably, $R_{xy}$ is seen to exhibit several developing quantized plateaus at $\pm\frac{h}{2e^2}, \pm\frac{h}{6e^2}, \pm\frac{h}{10e^2}, \pm\frac{h}{14e^2}$ for electrons (+ sign) and holes (- sign), where e is the elementary charge and h is the Planck constant. The half-integer QHE is an electronic hallmark of single-layer graphene[2,13,52], with vanishing $R_{xx}$ and quantized Hall plateaus occurring at the Landau Level (LL) filling factor $i = nh/eB = 4(N+1/2)$ (where n is the 2D carrier density, and N is a non-negative integer). The LL filling factor ($i$ = 2,6,10,14) for the observed quantum Hall states in Fig. 4(d) is indicated near the corresponding Hall plateaus. Observation of QHE is an important indication that the scalable CVD grown graphene film possesses the intrinsic graphene properties with electronic quality approaching or comparable with exfoliated graphene flake from graphite. The role of Si-SiO$_2$ substrate in the discovery of SLG is important, however it is not an ideal substrate to graphene. Scattering of charge carriers by charged impurities in SiO$_2$ is considered an important factor limiting the carrier mobility (typically ~$10^4$ cm$^2$/Vs or lower) in graphene on SiO$_2$[1,2]. Scattering of charge carriers by optical phonons of $SiO_2$ substrate further limits the theoretical mobility of graphene to ~200 000 $cm^2/Vs$[53]. Tremendous improvement in the mobility of graphene to ~100 000 $cm^2/Vs$ has been observed for graphene on h-BN, leading to much better electronic properties. For instance, RT ballistic transport at micrometer scale[7], fractional quantum Hall effect (FQHE)[54], and long distance spin transport[55]. This strategy has been adapted to CVD-grown graphene[56], but the development of large scale h-BN substrate technology is still in an early stage[57,58].

## 5. Applications of CVD grown graphene films

Metal catalytic CVD method is now being used to grow large-area polycrystalline graphene films with high uniformity, showing promise for many applications[59]. Compared to other large-scale graphene synthesis methods, CVD grown graphene on copper foil substrates has been shown to have electronic transport properties comparable to those of exfoliated graphene on a small scale device[32]. And the production cost is relatively low among other methods. Despite the fact that CVD grown graphene films may be less perfect (the presence of domain boundaries, defects, wrinkles, impurities and inclusions of thicker layers) compared to those of exfoliated graphene, such films (due to their large size and ability to be transferred to arbitrary substrates) would still facilitate applications in many areas, including flexible electronics[13,60], photonics devices[61-71], sensors and bio-applications[72-74]. Both academic laboratories and industries have demonstrated many devices in these aspects.

Synthetic graphene films produced by CVD method meet the electrical and optical requirements to substitute the indium tin oxide (ITO) traditionally used as a transparent conductive coating (TCC) in flexible electronics. Such graphene films allow a sheet resistance in a range of 50 to ~300 Ω/□ with a transmittance of ~90% compared that of a typical TCC. Additionally, graphene has ten times higher fracture strain compared to that of ITO[4]. Such graphene based TCC could be applied to touch screens, rollable e-papers, light emitting diodes (LED) and replacing ITO as the ubiquitous transparent conductor.

In addition to electronic applications, graphene also feature impressive optical properties arising from massless Dirac Fermions. Wavelength independent absorption (~2.3%) for normal visible light ($< $ ~3 eV)[5] and electrically tunable carrier transport properties[2] in graphene promise many electrically controllable photonic devices. A range of photonic devices using graphene have been demonstrated --- for instance, ultrafast graphene photodetectors[61,62], and ultrahigh gain graphene based photodetectors[63], optical modulators[64,65], mode-locked lasers[66], and graphene plasmonic devices[67-71], etc. Xia et al. demonstrated a graphene photodetector with optical bandwidth up to ~40 GHz[61]. However, further analysis suggests that the theoretical maximum bandwidth of a graphene photodetector can reach as high as 1.5 THz (in practice, 640 GHz limitation due to the capacitive (RC) delay) compared to that of InGaAs (150 GHz)[75] and Ge (80 GHz)[76]. Hence, the development of graphene photodetectors would be beneficial to the future high-speed data communications. Graphene is also found to be an intriguing plasmonic medium[68-70] recently as it provides the ability to control the plasmons (collective electron density oscillations that can be excited when light hits appropriate materials) in graphene by electrical voltage. These studies have shown that this wonder 2D material could be a useful component in future photonics.

Graphene is a promising material for sensing applications and bio-applications because graphene is highly sensitive to electrostatic perturbation by locally charged particles close to the surface[77]. By fabricating graphene FETs on a radiation absorbing substrate, such devices have been demonstrated to show the ability to detect electromagnetic radiations (light, x-ray, and γ-ray)[78,79]. The technical approach is to utilize the highly sensitive dependence of the electrical conductivity of graphene on local electric field, in which charge ionization is created when energetic radiation interacts the underlying radiation absorbing substrate. Also graphene is chemically inert and pure, and it can be functionalized by other molecules as a drug delivery vehicle. Moreover, the gas and liquid impermeability property of graphene[80,81] makes graphene a potential candidate in bio-compatible protective coatings or barrier films[73,74], which can be used, for example, in biomedical implants and devices. Before graphene can fulfill the requirements in the biomedical area, we have to understand its biocompatibility and acute and long-term toxicity under manufacturing and biological environments.

## 6. Conclusions

In summary, synthetic graphene grown by CVD on Cu foils has been found to be a promising way in producing large-scale, high quality, and uniform graphene films for graphene based applications. The electronic property of CVD grown graphene film is approaching that of exfoliated graphene. Other advantages of this method are relatively low production cost, large-scale and reproducible production compared to alternative graphene synthesis methods. Applications using CVD graphene films have been found in numerous fields, such as transparent conductive layers, nanoelectronics, flexible macroelectronics, photonics, sensors and bio-applications in spite of the imperfections found on CVD graphene. Since the development and current market of graphene applications are driven by the production and quality of this material, further improvements are desired the wide use of synthetic CVD graphene technology.

## Acknowledgements

We acknowledge partial support from NSF, NIST and DTRA for our synthetic graphene research.